# Direct integration of atomic precision advanced manufacturing into middle-of-line silicon fabrication

E. M. Anderson ; C. R. Allemang ; A. J. Leenheer ; S. W. Schmucker ; J. A. Ivie ;
D. M. Campbell ; W. Lepkowski; X. Gao ; P. Lu; C. Arose; T.-M. Lu ; C. Halsey; T. D. England ;
D. R. Ward ; D. A. Scrymgeour ; S. Misra

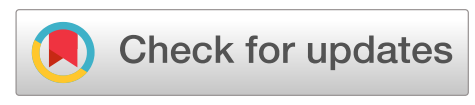

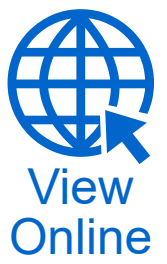

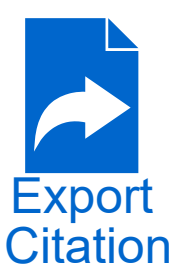

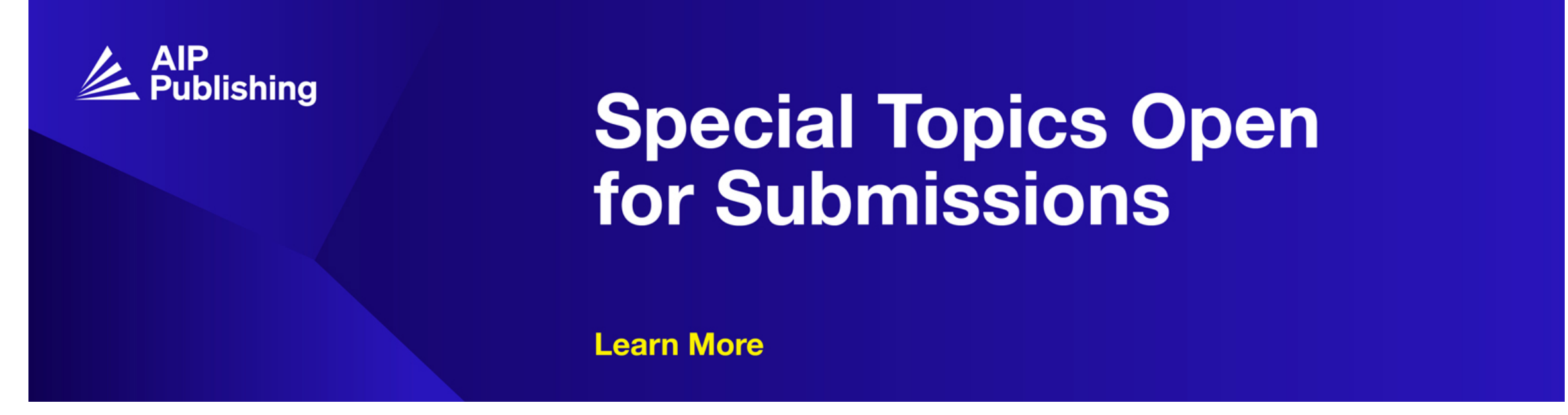

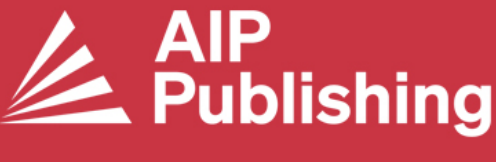

# Direct integration of atomic precision advanced manufacturing into middle-of-line silicon fabrication

E. M. Anderson, C. R. Allemang, A. J. Leenheer, S. W. Schmucker, J. A. Ivie, D. M. Campbell, W. Lepkowski, X. Gao, P. Lu, C. Arose, T.-M. Lu, C. Halsey, T. D. England, D. R. Ward, D. A. Scrymgeour, and S. Misra[a)]

## AFFILIATIONS

Sandia National Laboratories, Albuquerque, New Mexico 87185, USA

[a)]**Author to whom correspondence should be addressed:** smisra@sandia.gov

## ABSTRACT

Atomic precision advanced manufacturing (APAM) dopes silicon with enough carriers to change its electronic structure and can be used to create novel devices by defining metallic regions whose boundaries have single-atom abruptness. Incompatibility with the thermal and lithography process requirements for gated silicon transistor manufacturing have inhibited exploration of both how APAM can enhance CMOS performance and how transistor manufacturing steps can accelerate the discovery of new APAM device concepts. In this work, we introduce an APAM process that enables direct integration into the middle of a transistor manufacturing workflow. We show that a process that combines sputtering and annealing with a hardmask preserves a defining characteristic of APAM, a doping density far in excess of the solid solubility limit, while trading another, the atomic precision, for compatibility with manufacturing. The electrical characteristics of a chip combining a transistor with an APAM resistor show that the APAM module has only affected the transistor through the addition of a resistance and not by altering the transistor. This proof-of-concept demonstration also outlines the requirements and limitations of a unified APAM tool, which could be introduced into manufacturing environments, greatly expanding access to this technology and inspiring a new generation of devices with it.

## I. INTRODUCTION

Direct integration of new technologies into the manufacturing of complementary metal oxide semiconductor (CMOS) microsystems has the potential to enhance CMOS performance and accelerate the discovery of new device concepts by leveraging past investments in CMOS manufacturing. Realization requires identifying and addressing material and processing incompatibilities where the new technology can interfere with previously fabricated structures, and subsequent steps in manufacturing can alter the material produced by the new technology. Atomic precision advanced manufacturing (APAM)[1] provides unprecedented control over doping in silicon, ranging up to levels high enough to fundamentally change its electronic[2] and optical[3] properties and at a precision that scales down to single-atom placement, demonstrated in analog quantum simulators[4,5] and multi-qubit registers.[6,7] APAM is a strong candidate for direct integration because it uses silicon as a substrate[8] and provides control over the electronic structure and confinement in doped silicon that parallels germanium alloying in semiconducting silicon. Here, we formulate an APAM process module that inserts APAM-based features into Sandia National Laboratories' legacy $0.35\,\mu m$ CMOS manufacturing process, and a proof-of-concept that shows APAM can be directly integrated into CMOS manufacturing. Accomplishing APAM insertion into CMOS requires modifying the APAM process to accommodate the reduced thermal budget available in the middle of CMOS manufacturing and developing a new patterning technique compatible with manufacturing. Outside of the obvious benefit to APAM-related device research, this work provides a roadmap of how to integrate any material having a $\leq 600\,^{\circ}C$ thermal budget into CMOS manufacturing workflows more generally, which may be particularly relevant for modern technologies where access to manufacturing is limited.

APAM uses area-selective attachment of dopant precursor molecules to produce atomically abrupt and strongly doped planar shapes, referred to as delta layers. A simplified illustration of the process flow is shown in Fig. 1. A scanning tunneling microscope (STM) removes hydrogen from the cleaned and passivated Si(100) surface, where dopant precursors selectively attach.[9,10] These areas are then capped with epitaxial silicon to activate the dopants. This process has been demonstrated down to the single atom level[11] to produce spin qubits, where individual dopants are serving to confine electrons into a 0D structure. Many aspirational applications of APAM do not require single-atom templates but only high dopant densities. For example, the abrupt profiles produce confinement that can be useful in exploring quantum tunneling in devices like an areal tunnel field effect transistor, a low power switch that conducts when electrons confined to the buried 2D sheet can tunnel to a gated hole layer at the surface.[12,13] Other researchers' early attempts at producing three dimensional profiles have achieved an active density two to three times higher[14] than the state-of-the-art,[15,16] with near unity activation. This could mitigate the growing impact of contact resistance in scaled transistors. Both of these examples, one seeking to discover novel devices and the other to enhance conventional transistors, would benefit from direct integration with high-volume CMOS manufacturing.

Two limitations prevent the direct integration of APAM and CMOS, both of which we show simplified process flows for in Fig. 1. The first arises from their thermal incompatibility. At a high level, CMOS manufacturing generally orders processes such that the highest temperature steps occur first and later steps limit their maximum temperature and time to prevent alteration of existing structures. APAM typically uses a $\sim$ 1000 °C[9] vacuum anneal to remove the surface dielectric and produce an atomically ordered surface to template surface chemistry. However, the dopant profiles APAM produces begin diffusing above 500 °C,[17] implying that the earliest point APAM can be inserted is the very end of the front end of line (FEOL), in which transistors are formed. Unfortunately, subjecting the mostly formed CMOS transistors to such high temperatures changes their electrical behavior. Here, we identify a lower thermal budget process that produces atomically ordered surfaces and does not change the electrical behavior of the CMOS transistors. The second limitation arises from the incompatibility of hydrogen lithography with CMOS manufacturing. With hydrogen lithography, chemical precursors need to bind with reactive sites exposed by depassivation before they become contaminated, precluding performing photolithography and processing in separate tools. Instead of hydrogen lithography, we will leverage the polymer based resists used by the rest of CMOS manufacturing in the APAM process. An earlier effort circumvented both incompatibilities through heterogeneous integration (HI) of a chip with atomic-scale APAM devices and another with CMOS elements.[18] However, problems with alignment and interfacial defects preclude the ability to have a device span both wafers, so HI loses the ability to introduce APAM elements directly into CMOS transistors or accelerate the discovery of new devices that combine CMOS and APAM elements. Here, we will focus on direct integration precisely to keep both of these latter goals in focus.

In this work, we formulate an APAM process that addresses both the thermal and lithographic incompatibilities and then perform a proof-of-principle direct integration of APAM and Sandia's legacy 0.35 $\mu$m CMOS process. We start by pinpointing where in Sandia's CMOS process flow an APAM module can be inserted and determine a thermal budget which preserves the electrical characteristics of the transistors and of the APAM material in Sec. II. Next, we electrically characterize APAM-doped material produced by removing protective dielectrics from the surface of silicon using ion sputtering and annealing within this thermal budget to produce a crystalline surface in Sec. III. We then use this sputter and anneal process in combination with a dielectric hardmask to produce photolithography-compatible APAM-doped material with an activated density and resistivity comparable to the original process in Sec. IV. Individually, each of these steps builds on procedures well-known to the surface science and microfabrication communities they come from, but we combine them in Sec. V into a cohesive APAM module to produce devices using a flow that inserts the APAM module in between FEOL and back end of line (BEOL). We electrically characterize working CMOS transistors, APAM resistors, and a simple circuit where both work together, illustrated in Fig. 1(g). This work opens the door to the development of an integrated tool that performs the three key APAM steps—oxide removal, patterned doping, and silicon capping—enabling APAM in a cleanroom environment for more accessible device research and development.

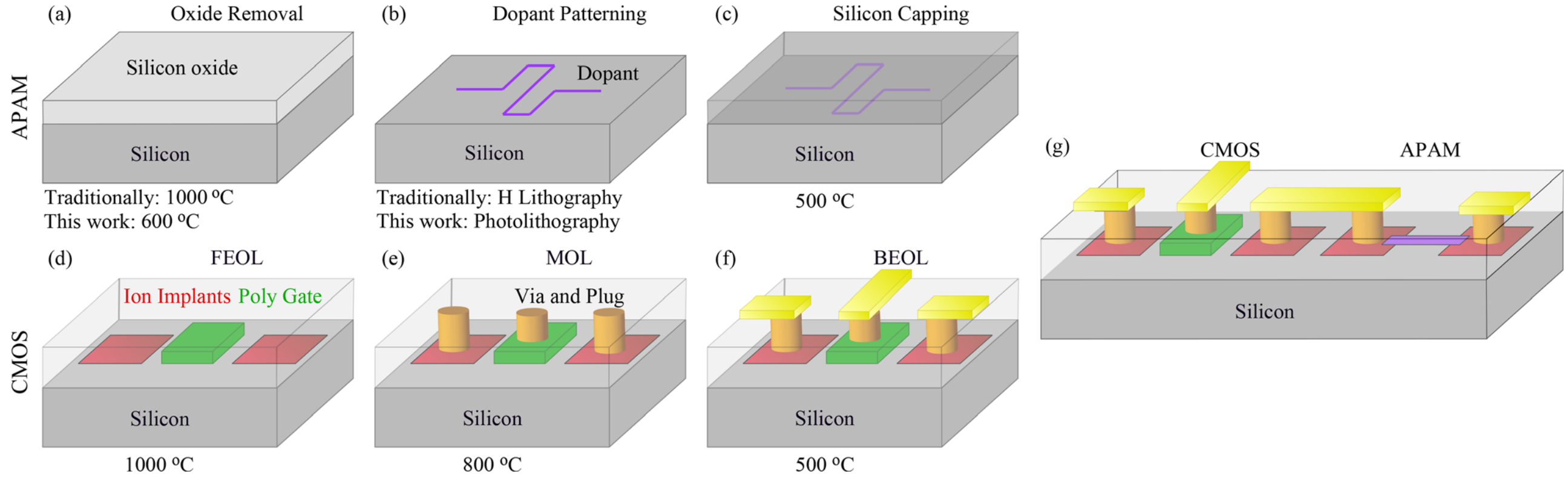

**FIG. 1.** The three basic parts of the APAM workflow include (a) removal of protective dielectrics, (b) patterned dopant incorporation (purple), and (c) silicon capping. (a) is typically accomplished with a high temperature anneal. (b) is typically accomplished using hydrogen lithography and exposure to a dopant precursor like phosphine. (c) is typically accomplished using low temperature molecular beam epitaxy. The three basic parts of a CMOS workflow include (d) front end of line (FEOL) for high temperature processing steps, (e) middle of line (MOL) for intermediate temperature processing steps, and (f) back end of line (BEOL) for low temperature processing steps. Sandia's legacy 0.35 $\mu$m CMOS process incorporates oxides and nitrides (clear), polysilicon gates (green), and ohmic implants (red) in (d), vias and plugs (orange) in (e), and metallization (yellow) in (f). (g) An integrated device with an APAM resistor attached to the drain of an NMOS transistor. The temperatures required for thermal compatibility are noted.

## II. APAM-CMOS THERMAL COMPATIBILITY

The CMOS flow can be divided into three stages: a high-temperature FEOL stage for transistor formation, an intermediate-temperature middle of line (MOL) stage for contact formation, and a lower-temperature BEOL stage to wire metallic interconnects [Figs. 1(d)–1(f)]. For the APAM delta layer to maintain its unique properties, the temperature of subsequent processing must be kept sufficiently low to avoid out-of-plane (vertical) diffusion of the delta layer after its formation. This will limit the earliest point in CMOS manufacturing where APAM can be incorporated. APAM material has historically been characterized using cryogenic Hall effect measurements, but the relationship of Hall data to layer thickness is indirect.[14,17,19–22] More recently, weak localization, seen in magnetotransport measurements at cryogenic temperatures, has been used to extract the electronic thickness of doped regions,[23,24] but focused on thermal conditions during epitaxy and not diffusion with post-epitaxial thermal treatment. Here, we extend those works specifically to study phosphorus diffusion in annealed delta layers after epitaxy. Figure 2(a) plots the delta layer width extracted from weak localization after heating samples in a rapid thermal annealer (RTA). A delta layer subjected to a 600 °C anneal still exhibits weak localization, a hallmark of 2D physics, with an extracted width that is nearly double that of an unannealed sample. In the rest of this work, we adopt 600 °C as the thermal budget available for the manufacturing steps following APAM.

Diffusion of the delta layer, thus, constrains the APAM insertion point to occur after the FEOL, which often concludes with dopant activation anneal and silicide formation steps that can involve thermal processing in the 800–1100 °C range. However, there are indications from previous works that CMOS transistors, which are mostly formed by the conclusion of FEOL, will change if subjected to flash annealing at 850–1250 °C, used by APAM to prepare a clean silicon surface. Previous work integrating micro-electromechanical system (MEMS) devices into legacy CMOS nodes (~ 0.25 $\mu$m) indicates that a few minutes at 600 °C preserves optimized dopant profiles at transistor junctions.[25,26] More recent work on stacked fully depleted silicon on insulator 14-nm devices indicates that several hours at 500 °C can be tolerated.[27]

To explore this thermal budget, we use Sandia's custom foundry process based on a 0.35 $\mu$m gate length silicon-on-insulator technology[28,29] as a demonstration platform (Sec. VII D). We start by performing extra anneals in an RTA at MOL after contacts to the transistors were made using tungsten plugs but before deposition of lower-temperature metals for routing [Fig. 1(e)]. Room temperature electrical testing of annealed p-channel (PMOS) and n-channel (NMOS) transistors shows degraded drain current vs gate bias ($I_D$-$V_{GS}$) curves, primarily from a shift of the threshold voltage (Sec. VII E). Figure 2(b) shows the threshold shift after a 20-minute RTA process at temperatures ranging from 400 to 850 °C, where the NMOS threshold shift exceeds our foundry's acceptable limit of ± 0.05 V when receiving thermal processing above 600 °C. PMOS devices show a smaller effect, where the threshold shift does not exceed acceptable limits even at 850 °C. Associated metrics, such as drive current shift accordingly, but we found no degradation of off-current. Shorter-duration anneals have lesser, but still measurable, effects. Possible contributors for the degradation include dopant uptake into the silicide[30,31] or development of fixed charge in the dielectrics near the channel.[32] Other annealing tests on transistors without tungsten plugs also show similar degradation of the NMOS threshold voltage (not shown), so the mechanism for threshold voltage shift in the 400–850 °C range is unlikely to result from dopant uptake by the silicide. For anneals above 850 °C, contact resistance degrades, and annealing at 950 °C visibly destroys the contacts by alloying the tungsten with the silicon. We estimate a viable MOL processing window of 500–600 °C based on these results. To allow insertion of APAM, we will need to reduce the thermal budget of the APAM surface clean.

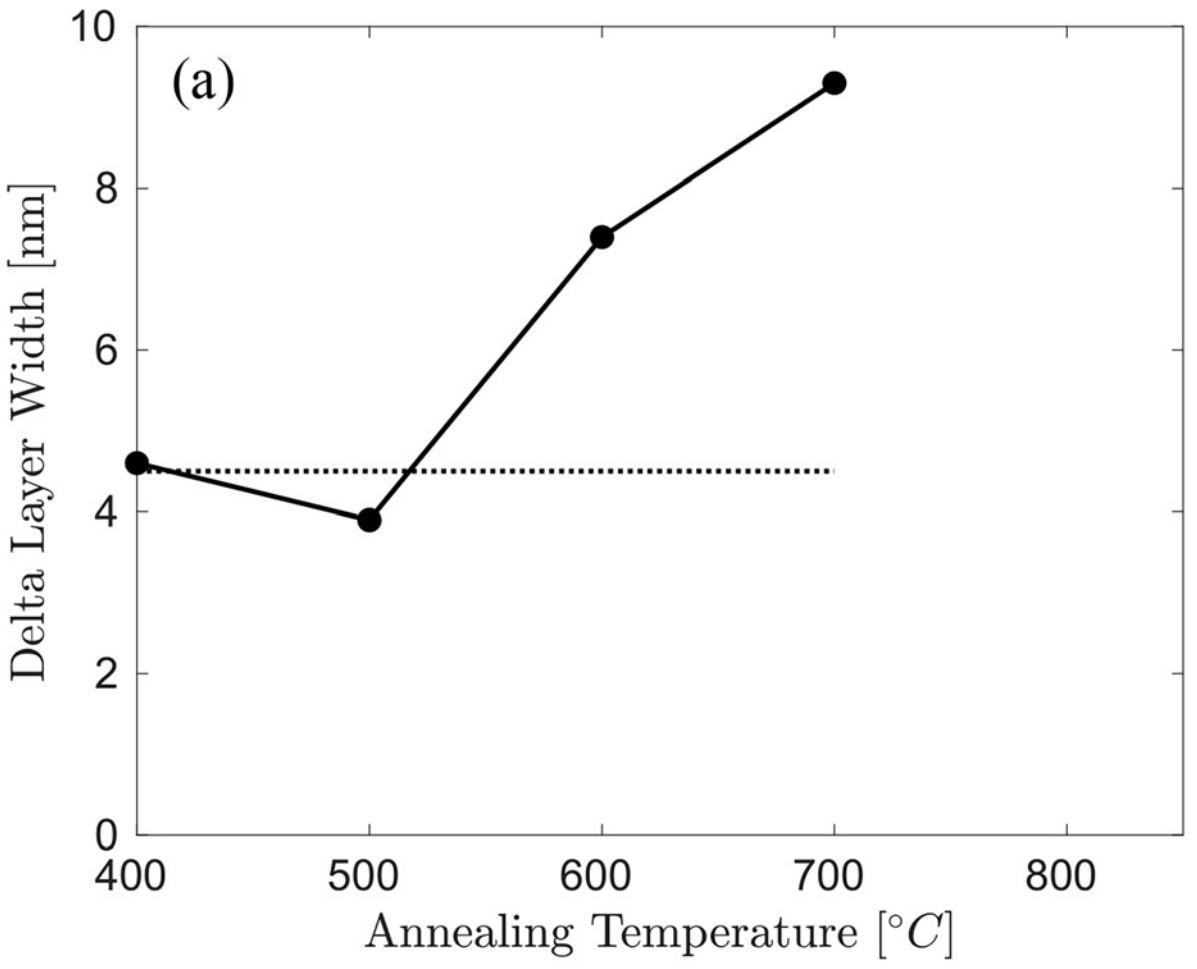

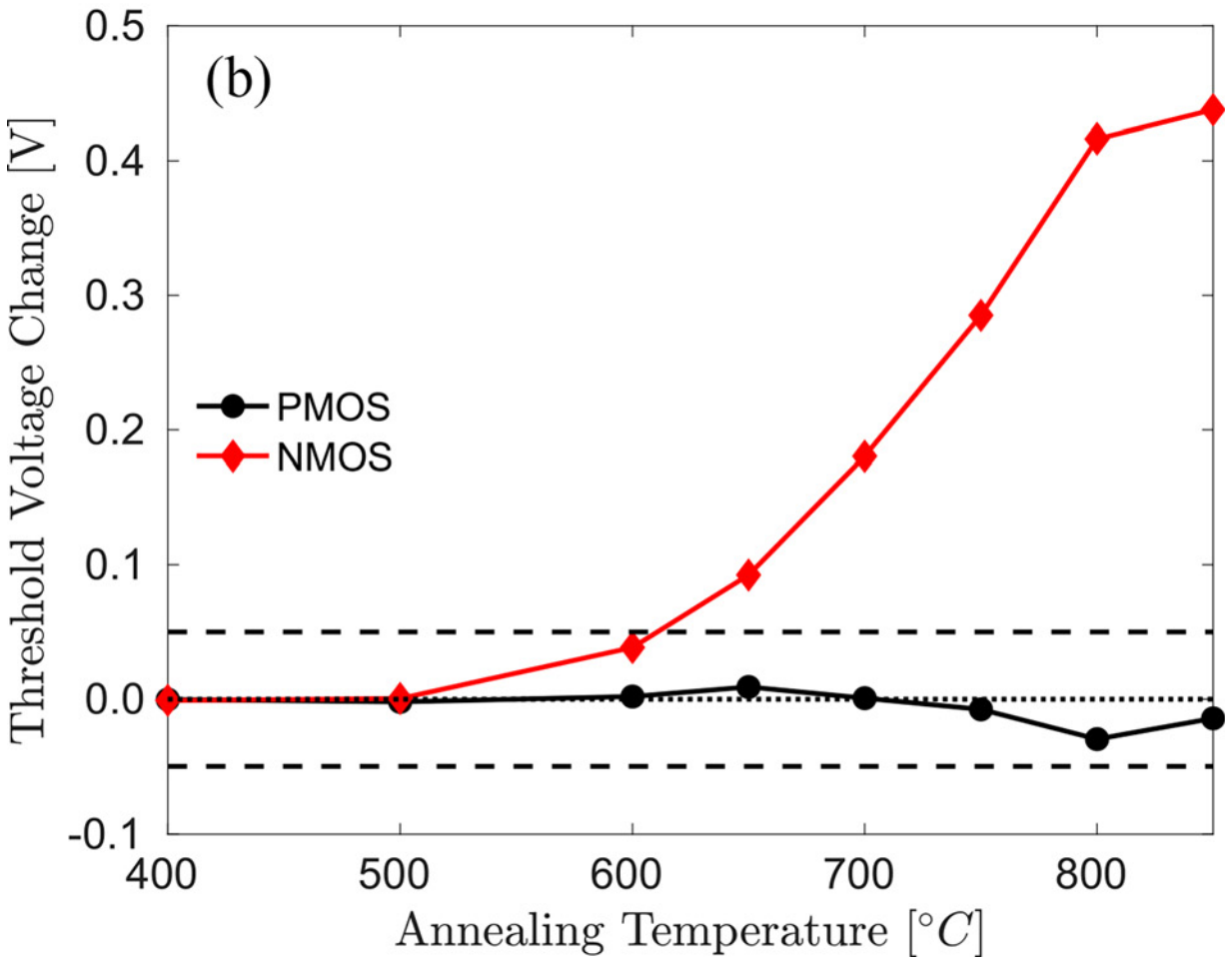

**FIG. 2.** Thermal budget tests for APAM delta layers and CMOS transistors. (a) Delta layer width extracted from weak localization for samples annealed 15 min at varied temperatures. (b) Threshold voltage change for transistors annealed 20 min at varied temperatures after contact formation. Dotted lines are guides to the eye, and the bounds for typical transistor threshold variations are shown as dashed lines. Each data point for both plots was taken by annealing a different sample, so some sample-to-sample variation in the relevant quantities is expected.

## III. APAM SAMPLE PREPARATION

Accommodating the limitations of the MOL of Sandia's 0.35 $\mu$m CMOS will require lowering the APAM thermal budget to $\leq 600\,^\circ$C. Fortunately, the only step of the typical APAM process that requires temperatures above 600 °C is removing the surface oxide through thermal desorption.[33,34] The surface preparation needs to produce a surface that is sufficiently crystalline to dissociate phosphine, canonically requiring three neighboring dimers of the $2 \times 1$ reconstruction to be present.[35] A popular choice for low-temperature oxide removal is chemical etching using aqueous hydrofluoric acid, but this produces a surface contaminated with carbon.[36] While many alternatives exist, including vapor hydrofluoric acid etching,[36] we chose noble gas ion sputtering as a proxy for dry etch tools common in industry. Combined with an anneal to recrystallize the amorphized surface, sputtering has been used to produce sufficiently flat and clean surfaces for STM imaging.[37] The recrystallization process, called solid phase epitaxy (SPE), proceeds at rates of 1 nm/min to tens of nm/min at temperatures between 500 °C and 600 °C for Si(100),[38] making it a good candidate given the available thermal budget. In SPE, silicon atoms at the amorphous-crystal boundary break bonds, diffuse a short distance, and remake bonds into a more ideal crystalline configuration.[38] This atomic rearrangement happens at a lower temperature than dopant diffusion in the bulk, which is driven by interstitial diffusion and then vacancy diffusion above 600 °C,[39,40] and might be an important factor in maintaining transistor electrical characteristics. A similar annealing process is already used in APAM device fabrication—the locking layer process used to minimize dopant segregation during epitaxy (Sec. VII B).[19,20] For the locking layer process, a thin silicon layer is deposited after doping the surface at ambient temperature, trapping most of the dopant atoms within this layer by limiting adatom-mediated diffusion during silicon deposition. A rapid anneal improves the crystallinity of the layer. Dopant diffusion is more likely to occur in this thin silicon layer during the anneal than in the bulk due to the motion of a high density of vacancies and interstitials formed during the low temperature epitaxy. However, as shown in Sec. II, after crystallization and additional growth, the thermal budget of the buried APAM layer resembles dopants in the bulk.

In Fig. 3, using high-angle annular dark field (HAADF) scanning transmission electron microscopy (STEM), we show the crystallinity of a sample sputtered for 30 min with neon ions accelerated at 2.0 keV at 60° from the surface normal, and then annealed at 550–600 °C for 15 s to complete SPE. The phosphorus delta layer is discernible in the cross-sectional STEM image,[41] and since it is incorporated on the surface, it marks the depth of the surface after sputtering and annealing in the cross section. A buried layer of defects is visible approximately 5 nm below the phosphorus delta layer, possibly from damage at the end-of-range for 2 keV neon ions.[42] This buried layer marks the extent of the observed damage caused by our sputter process, meaning that a 5 nm layer of silicon was amorphized and then recrystallized. SPE of amorphous silicon proceeds at 20–30 nm/min at 575 °C,[38] which is consistent with this 15 s anneal recrystallizing 5 nm of silicon. Further optimization through the use of lower energy ions can reduce the thickness of the amorphous layer and the amount of annealing needed to repair it. This will likely bring the buried defect layer to the surface, which may prove desirable if that can be healed through SPE, or undesirable if it cannot.

To establish a stronger correlation between crystallinity and electrical activation, we use STM to image the degree to which annealing heals the surface and compare these data to magnetotransport measurements after delta layers are made with these samples. These samples were sputtered with 1.5 keV neon ions, where the lower energy will produce roughly 75% of the amorphization depth in Fig. 3. Annealing the sputtered sample for $\leq 1$ min at 485 °C leaves the surface amorphous with no discernible atomic-scale ordering [Fig. 4(a)], which is consistent with the low SPE rate of $\sim 0.6$ nm/min.[38] Subsequently fabricated etch-defined Hall bars (Sec. VII C) were not conductive at 4 K. Increasing the annealing time at 485 °C to 15 min produces a rough, but crystalline, surface where the dimer rows of the $2 \times 1$ reconstruction are visible on small terraces, despite significant topographical variation [Fig. 4(b)]. Qualitatively similar results were obtained for 5 min at 500 °C, which is consistent with the higher SPE rate of $\sim 0.9$ nm/min at this temperature.[38] Etch-defined Hall bars made from these rough but crystalline surfaces all produce a carrier density in excess of $1.2 \times 10^{14}$ cm$^{-2}$ at 4 K. These results demonstrate that a high carrier density is correlated with the creation of a crystalline surface.

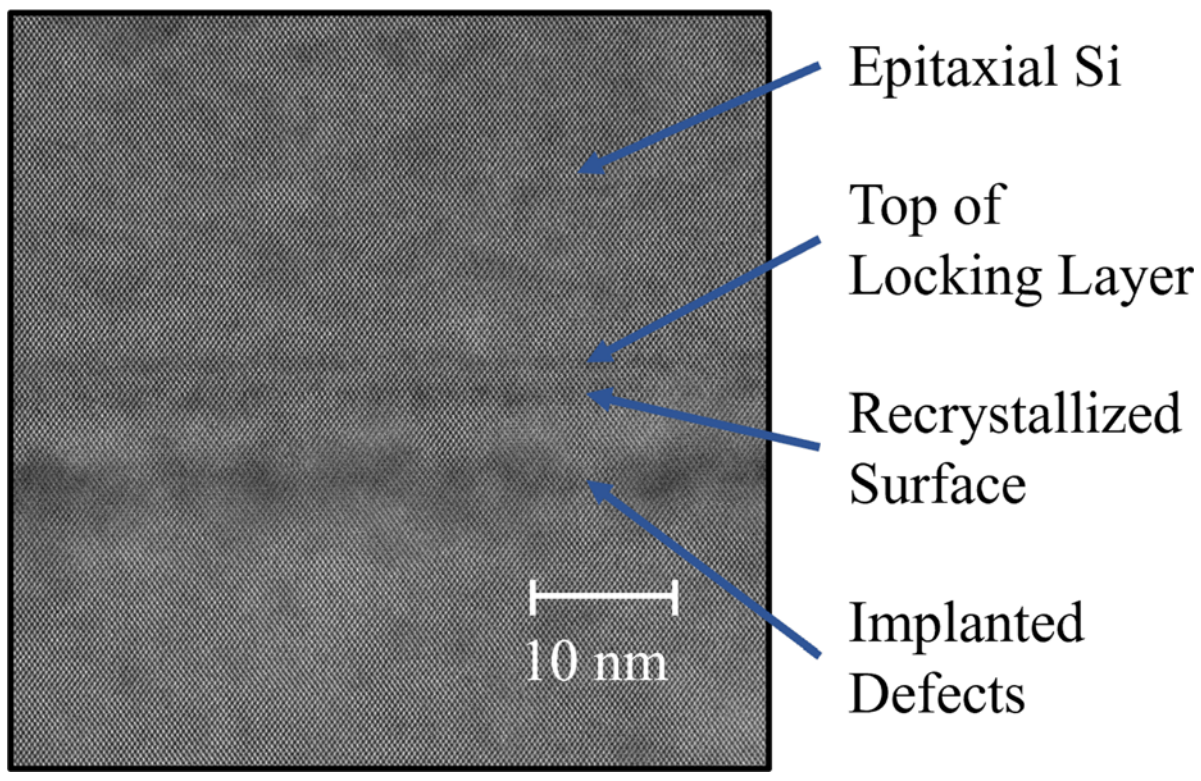

**FIG. 3.** STEM HAADF cross section of a sample that was sputtered with neon and annealed to recrystallize before completing the APAM process. Starting from the bottom of the image, a buried layer of defects is visible, followed by $\sim 5$ nm of recrystallized silicon, the phosphorus delta layer, the top of the locking layer, and the epitaxial silicon cap.

Going further to 560 °C and 600 °C for 15 min [Fig. 4(c)] not only produced a crystalline surface with larger terraces of dimer rows than lower temperatures but also revealed surface defects (e.g., in the upper-right corner of the image) that might arise from contaminants in the buried defect layer diffusing to the surface. For samples annealed for 15 min at different temperatures, the 2D carrier density plateaus after exceeding the threshold annealing temperature, where the surface becomes crystalline [Fig. 5(a)]. A surface where the dimer rows of the Si(100) $2 \times 1$ reconstruction are present is likely all that is needed for the decomposition pathways to proceed as proposed for various dopant precursors.[35,43,44] Conversely, an amorphous surface does not contain the ordered dimers used in these decomposition pathways, reducing dopant incorporation below the threshold for metallic conduction, $3.7 \times 10^{18}$ cm$^{-3}$ at 4 K.[45] The mobility continues to improve from 22 to 32 cm$^2$ V$^{-1}$ s$^{-1}$ and thus, the resistance continues to decrease, on increasing the annealing temperature from 485 °C to 600 °C [Fig. 5(b)]. This improvement in mobility correlates with a decrease in the root mean square (RMS) surface roughness of the layer from 274 pm to 167 pm, with the caveat that the smoothest layers examined here, ranging 123–132 pm, are amorphous and, thus, insulating at 4 K. To what extent this trend in mobility is purely a function of the roughness or is also influenced by the buried defect layer will require further work to determine. For comparison, our typical APAM process with a high-temperature surface clean produces material with similar densities ($1.5–2.0 \times 10^{14}$ cm$^{-2}$) and even higher mobilities (40–50 cm$^2$ V$^{-1}$s$^{-1}$), consistent with published results.[19] That material has an even lower RMS surface roughness (50 pm or less) and has no buried layer of defects from the sputter process.

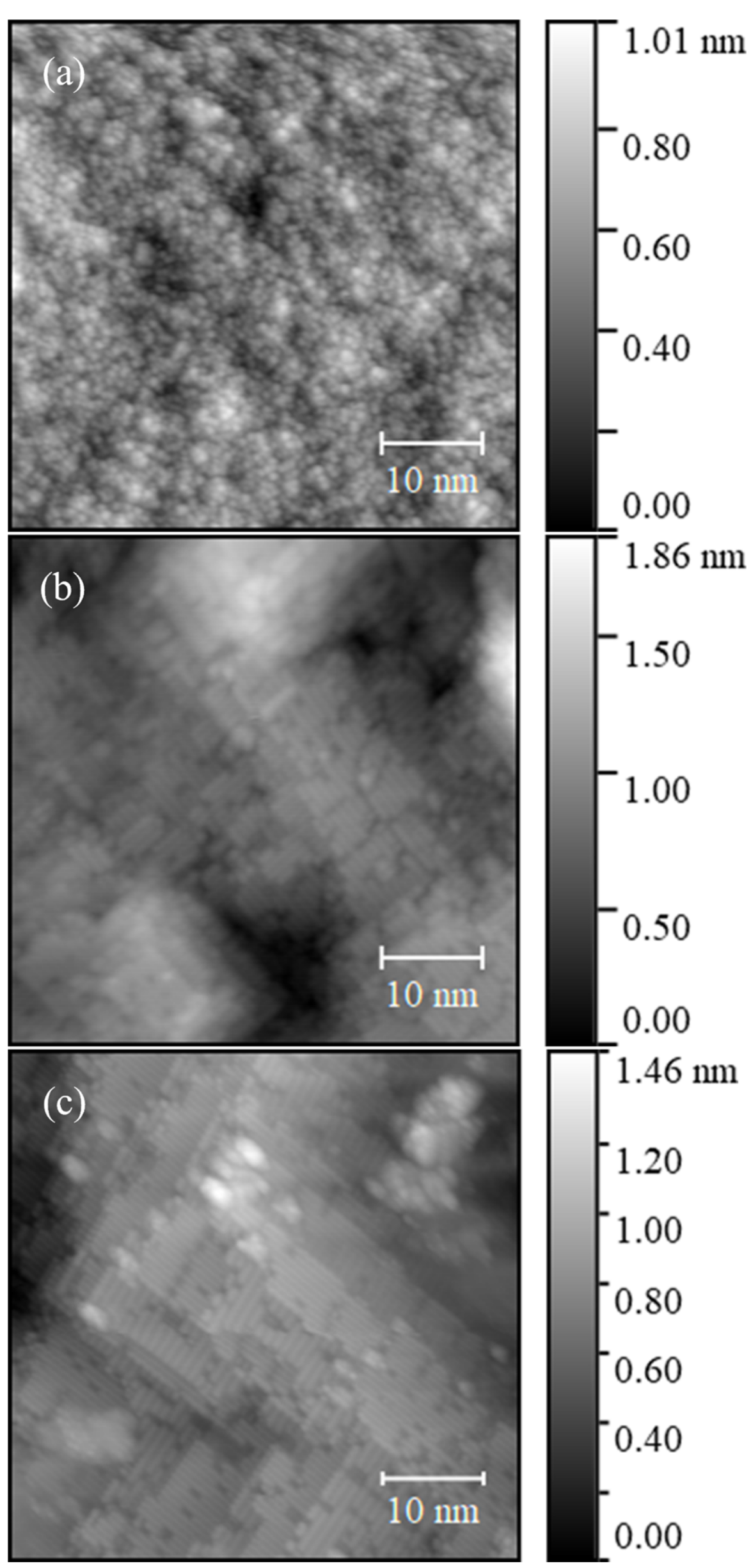

**FIG. 4.** STM topographical images at a sample bias of −2.0 to −2.5 V and tunneling current of 0.1 nA showing (a) a sample annealed for 15 s at 485 °C that remained amorphous without discernible atomic order, (b) a sample recrystallized after 15 min at 485 °C showing dimer rows of the 2 × 1 reconstruction on small terraces, and (c) a sample recrystallized after 15 min at 560 °C showing larger terraces of dimer rows and the accumulation of surface defects (e.g., in the upper right corner).

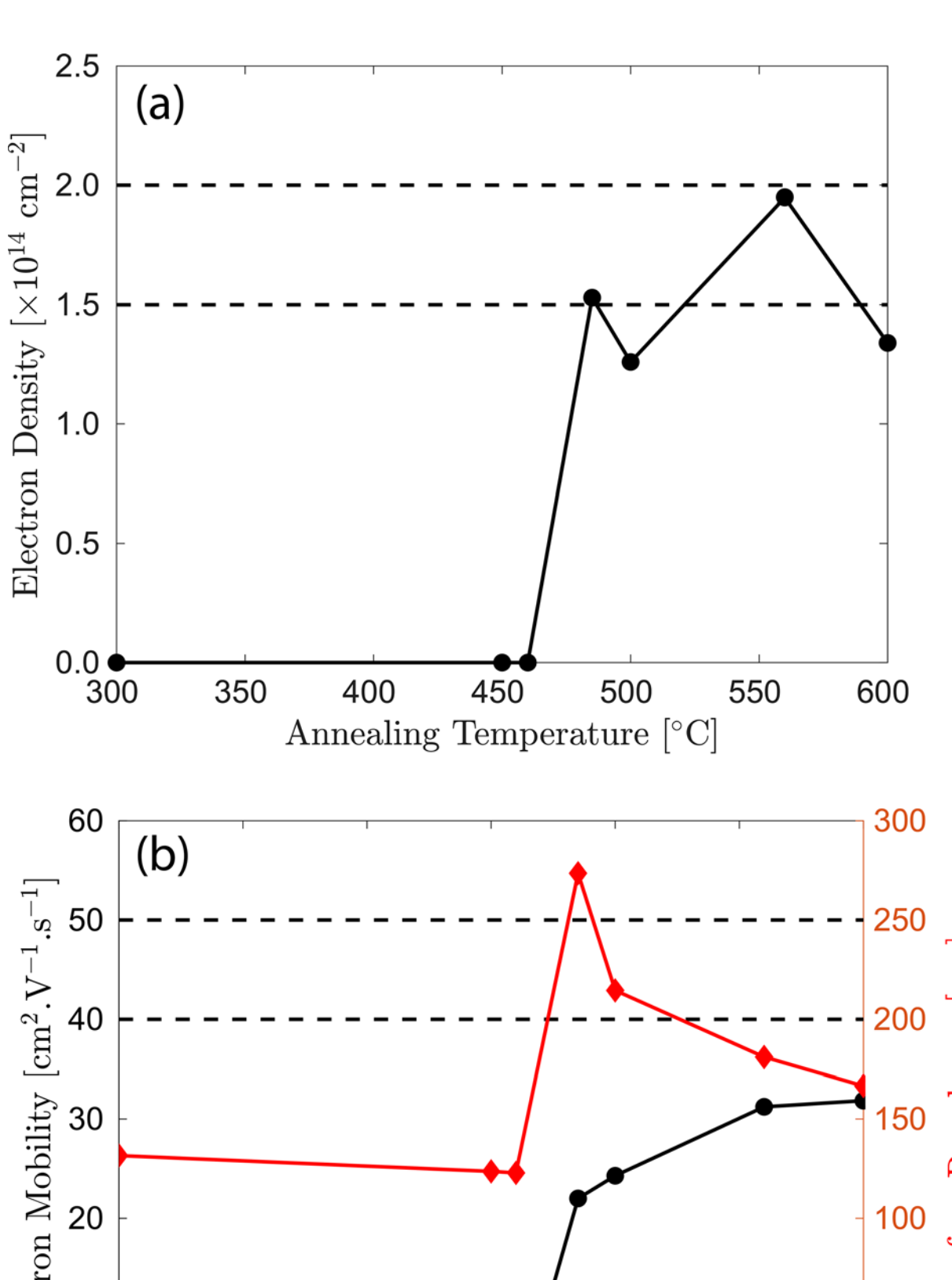

**FIG. 5.** Plots of 4K delta layer transport properties as a function of 15 min annealing temperature: (a) 2D active electron density and (b) electron mobility (black, left vertical axis) as well as RMS surface roughness (red, right vertical axis). Dashed lines on the plots indicate typical values for samples prepared by high temperature oxide removal.

This sputter and anneal process is a proofofconcept, using a toolset derived from the surface science community and is not optimized for manufacturing. Importantly, it achieves a density of activated dopants where changes to the electronic structure characteristic of APAM, including effects related to confinement, have been seen.[2] While the surface science and manufacturing toolsets are different, this sputtering demonstration provides an outline of an integrated APAM doping tool: the ability to remove dielectrics and produce a clean silicon surface with a modest thermal budget, incorporate dopant precursors, and encapsulate with epitaxial silicon. Notably, there is room for a manufacturing-based toolset to improve on the conductivity of the film shown here, which is important for applications, such as reduced transistor contact resistance. The amorphization and buried defect layer discussed above are particular to our choice of equipment and might not be a concern for manufacturing toolsets, which provide much greater control over the ion energy combined with a wider chemical palette. In particular, the Siconi process is well-established to remove oxide immediately before contact metallization without breaking vacuum,[46] and might provide a path toward cleaning the surface for APAM doping while limiting damage. Different surface cleans, such as atomic layer etching or reactive ion etching (RIE), might be feasible and are in wide use for other purposes. Similarly, the thermal anneal, which is a part of advanced contact modules, might provide a convenient way of healing the damage. Optimization of these processes might produce films with lower defect densities and having mobilities and densities closer to those produced by flash annealed samples, both almost double those seen here. Going further, a tool capable of quickly repeating the doping and encapsulation steps will be able to produce material having much higher conductivity by enabling three dimensional doping profiles. Overall, our proof-of-concept provides the thermal and crystallinity requirements for application-motivated optimization in the future. The final requirement for using such a toolset in CMOS manufacturing is a compatible lithography process, which is introduced in Sec. IV.

## IV. APAM LITHOGRAPHY

Selectively doping areas of the surface in APAM is typically accomplished using hydrogen lithography, but there is no simple way to incorporate hydrogen lithography into CMOS manufacturing. Hydrogen lithography proceeds by rendering the Si(100) surface chemically inert by attaching hydrogen to reactive dangling bonds and then selectively removing hydrogen in areas where dopant precursors will react with re-exposed dangling bonds. CMOS manufacturing relies on exposing different regions of a polymer film to either deep or extreme ultraviolet light, which, after reaction with a developer, will selectively mask the surface. In principle, the polymer resist can be replaced with hydrogen resist, and light can be used to inject enough energy into the Si-H bond to break it through either secondary electrons[47] or local heating[48] for deep ultraviolet light, or direct bond scission[49] for extreme ultraviolet light. However, the exposed dangling bonds will immediately react with the molecules in their environment and will, thus, form a native oxide while being carried from the photolithography station to an (aspirational) APAM processing station. While this can be circumvented by designing a photolithography stepper which works in an ultrahigh vacuum environment and where samples can be vacuum-transferred to an APAM process chamber, the complexity and cost of such a tool make this quite risky.

We forego hydrogen lithography and explore how to make APAM processing compatible with the polymer resist-based photolithography ubiquitous in CMOS manufacturing. The previously discussed reduced thermal budget of the sputter and anneal process will severely degrade polymer resist. Instead, we employ a common technique in CMOS manufacturing to transfer the pattern from the polymer resist mask to an oxide hard mask. A mask with areas of thick and thin dielectrics is formed by growing a thick dielectric on silicon, then using a polymer resist mask to selectively etch down to the silicon, and finally growing a uniform thin oxide. The sputter and anneal process, illustrated in Fig. 6, relies on the mask having two thicknesses of dielectric, and the sputter process removing a uniform amount of material. This exposes silicon only in the areas with the thinner dielectric. Dopant precursor molecules are only incorporated over areas where bare silicon is exposed, and while the silicon cap is deposited everywhere, it can be removed in post-processing outside of the thinner dielectric region. This is a key improvement over the traditional hydrogen resist processes used in APAM processing, as it provides scalability beyond serial scanning probe lithography and compatibility with standard parallel CMOS processes, such as photolithography, growth, and etch.

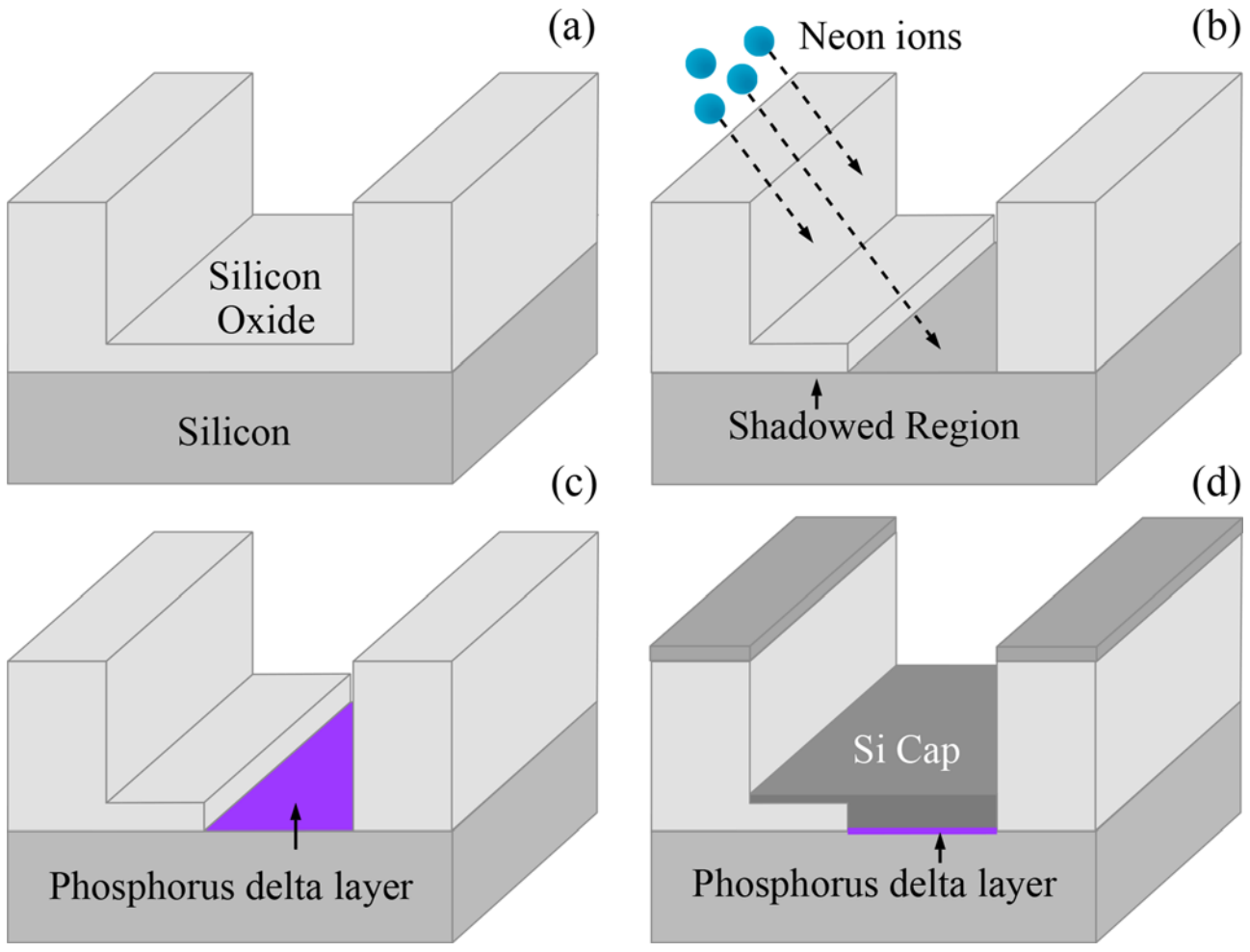

**FIG. 6.** Neon ion bombardment for surface cleaning an oxide hardmask, including the effects of shadowing on APAM doping. Shadowing will have the effect of doped areas being smaller in practice than is intended by the pattern in the hardmask. We use the terms corrected and uncorrected geometry for the actual doped area and the intended pattern area, respectively. (a) The photolithographically defined hardmask. (b) Neon ion bombardment to remove the thin oxide with a residual oxide in the device area as a result of shadowing from the hardmask. (c) APAM doping with phosphorus in the sputtered region. (d) Capping with epitaxial silicon.

To explore the ability of a hardmask to define the geometry of doped regions, we study cryogenic magnetotransport data from Hall bars having different widths. Sputtering is a physical process that uses ballistic neon ions and will only strip the thin oxide from part of the device area, with some shadowing from the thick dielectric mask on the side of the pattern nearest the sputter gun [shown in Fig. 6(b)]. Sputtering can also produce redeposition from sidewall sputtering of the thick dielectric mask on the far side. We first examine data from Hall bars defined using a 200 nm thick oxide mask in Fig. 7 using black symbols. We find that these Hall bars have an electron density lower than (down from 1.5 to $0.94 \times 10^{14}$ cm$^{-2}$) those in Sec. III, which were created by using a sputter-and-anneal process to produce a blanket layer of doped material and then etch-defining a Hall bar. The lower density is likely the result of a lower coverage of dimers at the surface, which are needed to decompose phosphine in a way that produces phosphorus well-bonded into crystalline silicon. This could arise from a higher density of surface defects from redeposition or from a dirtier plasma-enhanced chemical vapor deposition (PECVD) oxide used here, compared to the cleaner thermal oxide used in Sec. III. Surprisingly, we find the Hall bars here have a higher mobility (up from 32 to 42 cm$^2$V$^{-1}$s$^{-1}$) than those from Sec. III. A possible explanation is the 1.5 keV neon ions used in Sec. III, which produce a shallower defect layer and increased carrier scattering. The samples here used higher energy 2 keV neon ions, which would move the buried

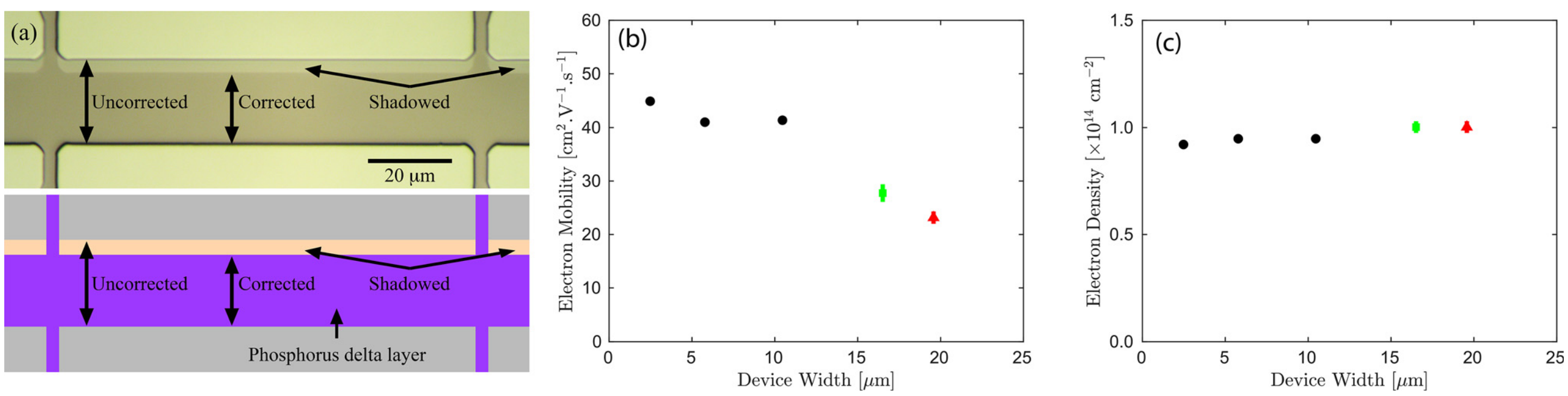

**FIG. 7.** (a) Optical image (top) and diagram (bottom) of the shadow from a 1 $\mu$m thick hardmask, showing the intended uncorrected width and the observed corrected width used for analysis of magnetotransport data. (b) Electron mobility and (c) density data from Hall bars patterned using a 200 nm thick hardmask is shown in black circles. Data from a 1 $\mu$m thick hardmask are plotted assuming the intended uncorrected geometry (red triangles) and the optically observed corrected geometry (green squares).

defect layer further from the phosphorus delta layer, decreasing carrier scattering.

The magnetotransport data from a Hall bar defined using a thicker 1 $\mu$m composite dielectric, similar to our CMOS chips, shows both visible and electrical signatures of shadowing. An optical image, Fig. 7(a), shows discoloration extending $\sim 3\,\mu$m from the hardmask where oxide was never removed, and dopants were never incorporated. The size of this shadowed region is nearly double that expected from the hardmask thickness of 1 $\mu$m and a sputter angle of 60° but is also clearly visible from the optical image. The magnetotransport data were first analyzed using the assumed geometry, where this shadowing is not accounted for [uncorrected width in Fig. 7(a)] and is shown as red triangles in Figs. 7(b) and 7(c). Shadowing was seen to shrink the width of the Hall bar from 20 $\mu$m to 17 $\mu$m, and increase the number of squares by nearly 20%. Analyzing the magnetotransport data using this corrected width [Fig. 7(a)] yielded the green squares in Figs. 7(b) and 7(c), with negligible impact on the electron density, but a 25% increase in mobility compared to the uncorrected width. The thick hardmask sample has a carrier density similar to the thin hardmask samples, both less than the samples from Sec. III. Once more, this might arise either from redeposition from the hardmask wall during sputtering, or the use of a dirtier PECVD oxide instead of a cleaner thermal oxide. Conversely, these thick hardmask samples have similar mobility to the samples in Sec. III, both lower than the thin hardmask sample. This might come from both being prepared using 1.5 keV neon ions and having a shallower defect layer than produced by 2.0 keV ions in the case of the thin hardmask sample. A shallow defect layer might scatter carriers more, reducing mobility.

## V. CMOS + APAM

Throughout this work, we have defined the thermal budget of Sandia's 0.35 $\mu$m CMOS process (Sec. II), developed a modified APAM process to reduce the processing temperature for compatibility with the CMOS thermal budget (Sec. III), and established a patterning technique to integrate the modified APAM process with CMOS (Sec. IV). In this section, we address the open question of whether APAM can be directly integrated with CMOS technology through three key demonstrations: (1) An APAM resistor [Fig. 8(a)], a strip of phosphorous-doped silicon formed and patterned using our modified APAM process, shows decreased resistance compared to an APAM cell without delta doping, indicating success of our combined process and patterning; (2) the CMOS component on the same chip remains operational throughout the integration process, confirming APAM and CMOS compatibility; and (3) we can connect the APAM resistor from point (1) in series with the drain of the NMOS in point (2) using on-chip wiring, allowing us to measure the expected electrical behavior.

For this demonstration, we fabricated a CMOS die we refer to as the CMOS test platform (Sec. VII D), which includes discrete process monitors and integrated transistors, using Sandia's 0.35 $\mu$m process. These custom-designed chips were specifically tailored for a split fabrication process, where APAM is inserted at the MOL, to ensure compatibility with our modified APAM process and patterning and post-APAM metallization in a separate cleanroom environment, thereby establishing a proxy split fabrication process. As illustrated in Fig. 8(c), wafers were pulled from the SiFab within the Microsystems Engineering, Science and Applications (MESA) facility at Sandia, after the FEOL and contact formation. The wafer was then diced into die of $5 \times 9$ mm that contained the full CMOS test platform while meeting the size restrictions of the tool used for our modified APAM process. Select pieces from the wafer then received a modified APAM process, developed in Sec. IV, at the MOL to facilitate integration. The MOL processing focused on a special-purpose cell that was included to facilitate the creation of an APAM resistor connecting two implanted regions, which we term an APAM cell. Within an APAM cell, a window was opened in the oxide + nitride hardmask using RIE to expose the silicon above and between the implants [Fig. 8(d)]. The APAM sputter processing was performed [Fig. 8(e)], then the sample was optionally dosed with phosphine for doping, followed by epitaxy of a silicon cap [Fig. 8(f)]. The epitaxial silicon was a blanket deposition across a chip and was removed using photolithography and a plasma etch, leaving it only within the vicinity of the APAM cell. Finally, the BEOL metallization was performed in a separate Sandia facility [Fig. 8(g)].

With the test platform established, we proceeded to validate that the modified APAM process and patterning remain effective when integrated with Sandia's 0.35 $\mu$m CMOS process. In Fig. 9(a), we present the resistance of APAM cell windows, having at least partial openings, without doping, and with delta doping as a function of measurement temperature. The delta-doped sample exhibits a room temperature resistance of approximately 11.5 $k\Omega$. With the geometry of the implants contacting the APAM doping and the geometry of the APAM cell window, we expect the resistor to be between 3 and 1 squares, yielding a sheet resistance of $3.83\,\mathrm{k}\,\Omega/\mathrm{sq}$ to $11.5\,\mathrm{k}\,\Omega/\mathrm{sq}$. We note that this sheet resistance is greater than the $1/ne\mu = 2.2\,\mathrm{k}\,\Omega/\mathrm{sq}$ (where $n$ is the carrier density, $e$ is the electron charge, and $\mu$ is the mobility) expected from thick hardmask-defined samples measured at cryogenic temperatures in Sec. IV. Three factors might contribute to this discrepancy. First, the resistivity is expected to rise between cryogenic and room temperature. Flash cleaned samples show a factor of two change in resistivity from cryogenic temperatures to room temperature.[50] Unfortunately, leakage currents become significant in our Hall bar samples at elevated temperatures, precluding us from measuring them at room temperature. Second, this measurement was a two point measurement and includes contributions from parasitic and series resistance components (e.g., contact resistance), while the previous were four point measurements that factor out series resistance. Finally, it is equally likely that there are further issues with patterning and dosing the small APAM cell window, as discussed in Sec. IV, leading to other electrical effects. For example, as temperature decreases, the current-voltage curves become non-linear at around 100 K, possibly indicating the contacts are a source of a temperature-dependent parasitic series resistance due to a Schottky junction. In all cases, the resistance can likely be improved through process optimizations, including repeating the doping and capping process to produce multilayers.[14]

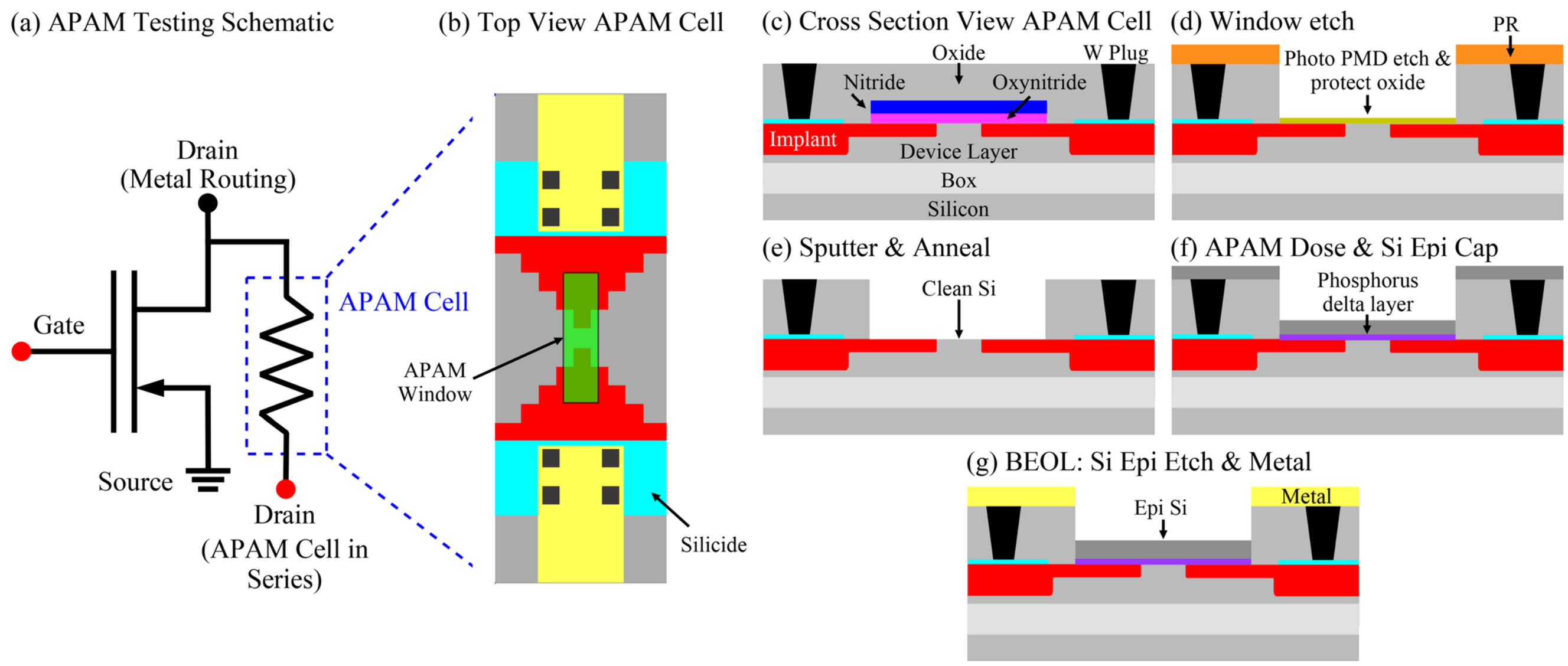

**FIG. 8.** (a) A schematic of an APAM cell integrated with a CMOS circuit acting as an optional resistor connected in series with the drain of an NMOS device. (b) Top view schematic of the APAM cell. Schematic cross sections showing the process used to selectively open up regions of silicon for APAM processing: (c) The chip after FEOL through implants, dielectrics, vias, silicides, and tungsten plugs. (d) Photoresist to define the dry etch for the APAM window opening and subsequent thin PECVD oxide deposition to protect the silicon prior to APAM. (e) Neon sputtering to remove the thin oxide and recrystallization of the silicon surface. (f) APAM doping with phosphorus and capping with epitaxial silicon. (g) BEOL to etch away the epitaxial silicon outside of the APAM window and patterning and deposition of metal routing.

To confirm the impact of APAM doping, we compare an APAM cell with and without doping and investigate the resistance of both as a function of temperature [Fig. 9(a)]. The room-temperature resistance of the undoped APAM cell is nearly five orders of magnitude greater than the doped cell, indicating that APAM doping significantly reduced resistance. Furthermore, the resistance of the undoped APAM cell rises sharply as temperature decreases, while the doped cell remains relatively flat, indicating doping near or above the metal to insulator transition. The latter is consistent with the relatively modest temperature dependence shown earlier in four-point measurements.[50] The undoped APAM cell forms two back-to-back pn junctions, between the n-type implants, the p-type device layer, and the unintentionally doped p-type APAM epitaxial cap.[34] In this case, there will always be a reverse biased pn junction leading to low current. Here, we make an important note that design considerations are needed to avoid making a forward biased pn junction between a doped APAM cell and the substrate, and to keep the resistance of the APAM cell relative to other device terminals in mind during circuit design. For example, if an undoped APAM cell is in series with the gate of a transistor device, unpredictable transistor and circuit behavior will result from having a gate that is nearly an open circuit. Finally, we show that the APAM cell does not work when part of the processing fails. In Sec. IV, we demonstrated that shadowing should be a consideration when using this process for APAM patterning. We measured an APAM cell that was completely shadowed (i.e., the APAM cell has no exposed silicon) at room temperature and observe a similar resistance to the undoped APAM cell [Fig. 9(a)].

Beyond confirming that APAM can be incorporated into a CMOS chip, we demonstrate that the CMOS components remain fully functional. First, in Fig. 9(b), we present the transfer characteristics for a control NMOS device with and without APAM processing. The APAM process may result in a slight increase in turn-on voltage and decrease in output current. This demonstrates that the CMOS chip can withstand the modified APAM processing and patterning. We then take an NMOS device and integrate the characterized APAM wire using on-chip wiring [Figs. 8(a) and 8(b)]. We note that in Figs. 9(c) and 9(d), the APAM resistor was connected in series with the drain for all measurements; however, bond pads provided direct access to the drain (black dot—Metal Routing) or connection through the APAM in series with the drain (red dot—APAM wire in Series w/ Drain). The transfer characteristics after integration wiring are shown in Fig. 9(c). From measurements taken using the Metal Routing bond pad (black line), we can extract a threshold voltage, taken to be the $x$-axis intercept of the linear region, of 1.62 V. Furthermore, we extract a subthreshold swing, taken to be the average inverse of logarithmic rate of device turn on, of 89.2 mV dec$^{-1}$. The threshold voltage and subthreshold swing confirm enhancement-mode behavior and effective electrostatic control of the channel at the oxide interface, respectively. The output characteristics, shown as symbols in Fig. 9(d), have linear behavior at low drain bias before saturating at higher bias. Linear behavior at low drain bias confirms that the contacts to the NMOS device remain ohmic.

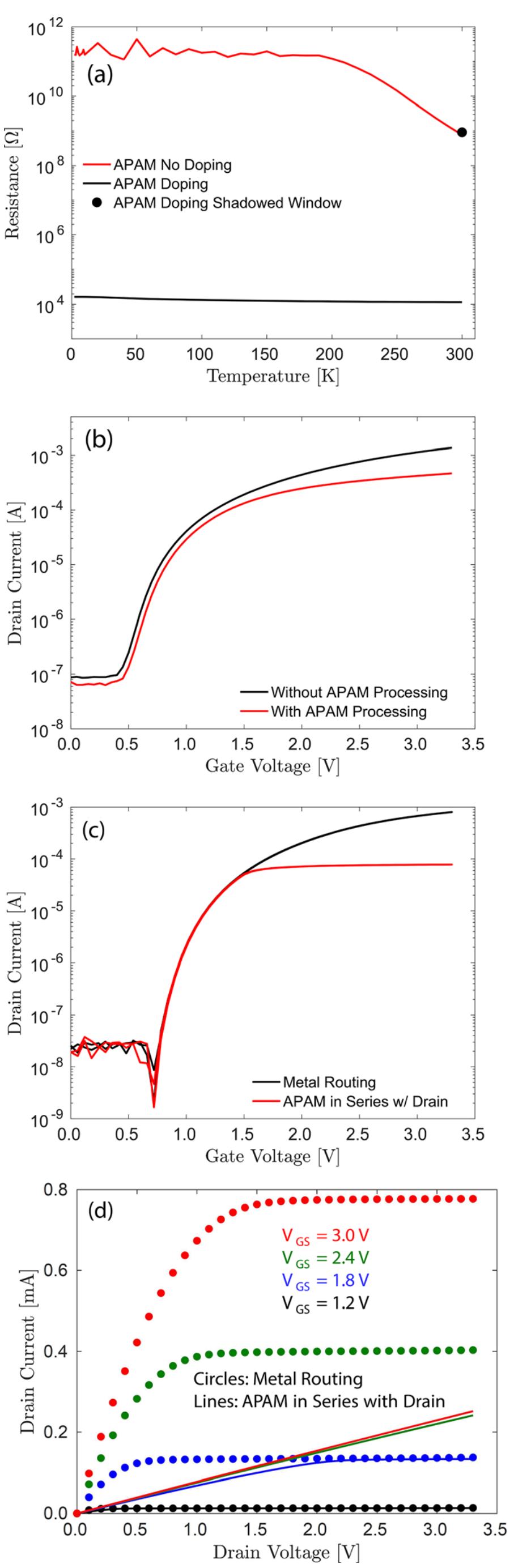

**FIG. 9.** (a) Resistance of an APAM cell resistor receiving all process steps without (red line) and with (black line) dopant dosing as a function of measurement temperature. A single point shows the room temperature resistance of a shadowed APAM window (black circle). (b) Transfer characteristics ($I_D - V_{GS}$) of a control NMOS without (black) and with (red) APAM processing. (c) Transfer characteristics ($I_D - V_{GS}$) and (d) output characteristics ($I_D - V_{DS}$) of the same NMOS device measured under two conditions: only metal routing (circles) and with an APAM resistor routed on-chip in series with the drain (lines). (b), (c), and (d) were taken at room temperature with (b) $V_{DS} = 3.3$ V, (c) $V_{DS} = 1$ V, and (d) $V_{GS}$ from 1.2 to 3 V in steps of 0.6 V.

Finally, we characterize the NMOS with the APAM resistor connected in series with the drain using the APAM in Series with Drain bond pad (red dot). The turn-on behavior in Fig. 9(c), red line, remains consistent for both, with the only change being a flattening of the drain current at a gate bias of $\sim 1.5$ V with the APAM resistor. This maximum drain current indicates when the resistance of the APAM resistor exceeds the channel resistance. We can confirm this by taking the slope of the output characteristics at a gate bias of 3 V [Fig. 9(d), red line], which gives a resistance of approximately 13 k Ω. The majority of the 13 k Ω can be accounted for by the APAM resistor, with the rest coming from the NMOS and associated parasitics. At a lower gate bias of 1.8 V (blue line), the drain current goes from being limited by the APAM resistor to being limited by saturation in the channel above 2 V. Finally, at an even lower gate bias of 1.2 V, the drain current is limited by saturation in the channel for much of the range of drain bias. We acknowledge that the proof-of-concept demonstrated here had low yield of functional CMOS + APAM devices, primarily due to the shadow masking discussed above. We anticipate that adapting this hardmask approach to industry-standard tools for lithography, etch, epitaxy, and ohmic contacts would ultimately reach device yields consistent with those of the foundry implementing the APAM doping after sufficient process development. This successful integration and understanding highlight the compatibility of our modified APAM process with CMOS to exploit CMOS + APAM for future microelectronic applications.

## VI. CONCLUSION

In this manuscript, we have formulated an APAM module, which achieves both thermal and lithographic compatibility with Sandia's legacy 0.35 $\mu$m CMOS process in between FEOL and BEOL, sacrificing single-atom placement but preserving high dopant density. The module starts with defining a hardmask after FEOL, as detailed in Sec. IV. This is followed by four steps in ultrahigh vacuum from Sec. III: sputtering to expose clean silicon in the thin parts of the dielectric mask, annealing to crystallize the surface, selectively incorporating a dopant precursor, and capping with low-temperature silicon by molecular-beam epitaxy (MBE). Finally, a microfabrication step removes the capping silicon from the thick part of the dielectric mask. This is a proof-of-principle direct integration that makes different process choices than those taken historically for both the surface preparation and area-selective dopant incorporation steps in APAM, replacing thermal oxide desorption and hydrogen lithography, respectively.

The significance of this proof-of-principle integration of APAM and CMOS is both conceptual and practical. The first advance is the articulation of a detailed methodology for direct integration of a boutique process. This is often too risky when existing microfabrication tools cannot recreate the desired process in the manufacturing facility. The inability to access the details of proprietary CMOS workflows can also make ambitious direct integration inaccessible to researchers. The roadmap used here can likely be extended to other boutique processes that have similar thermal envelopes, and are being integrated with other manufacturing workflows, including modern ones. The second advance is the specification of a tool capable of bringing APAM out of the laboratory and into the microfabrication facility, expanding access from a limited number of domain experts to a much wider community. In practice, we anticipate that a more optimized process will be needed by industry, given access to technologies which are common in manufacturing but are inaccessible at the laboratory scale, including deep and extreme ultraviolet photolithography, proprietary low-temperature dielectric etches, ultrafast rapid thermal annealing, and low-temperature chemical vapor deposition. The high throughput a manufacturing process enables is a prerequisite for yield exploration or optimization, and is unlikely to be related to the bespoke process detailed here. Conversely, an integrated tool that can perform sputter and anneal, chemical doping, and silicon homoepitaxy is simple and flexible enough to incorporate into university-scale clean rooms to spur innovation. Finally, the potential impact of APAM-CMOS integration can best be imagined by looking at the design space for semiconductor electronic structure enabled by alloying silicon with germanium, and the large number of devices this has inspired. We envision that the wider availability of APAM will result in the device community no longer thinking of doping as just adding either electron or hole carriers to the parent semiconducting material but a doped material having novel properties itself. Only then can the design space for doped material serve as a source of inspiration for the discovery of new devices.

## VII. EXPERIMENTAL METHODS

### A. Starting material

The material used to make Hall bars in Secs. II, III, and IV was p-type Si(100) substrates with 1–10 Ω cm resistivity, processed at the wafer level starting with two rounds of SC1, SC2, dilute HF cleans, and a 3 nm dry thermal gate oxidation to protect the silicon surface. Standard photolithography followed by a $CF_4$/Ar inductively coupled plasma (ICP) etch created alignment marks for all subsequent process steps.

For the blanket-processed samples in Secs. II and III, the sacrificial oxide was then stripped in a 6:1 buffered oxide etchant (BOE), and a final thin 3 nm thermal oxide was grown to protect the surface before dicing into individual die. The oxide hardmask samples in Sec. IV underwent solvent cleaning and oxygen plasma prior to plasma-enhanced chemical vapor deposition (PECVD) of a thick 200 nm layer of $SiO_2$ at 250 °C. Hall bars were patterned into the field oxide using standard photolithography and etched in 6:1 BOE. Samples were then sent for PECVD to produce a 3 nm thin $SiO_2$ layer inside the patterned areas. Hardmask samples with a dielectric stack matching the CMOS parts underwent a similar process to the oxide hardmask samples, where the thick oxide was replaced by a dielectric stack and a dry etch was performed as described in Sec. VII D. FEOL fabrication of CMOS chips prior to APAM integration is described in Sec. VII D.

### B. APAM processing

Before loading into an ultrahigh vacuum (UHV) system, all APAM samples were cleaned in ultrasonic baths of acetone and isopropanol, followed by a radio frequency oxygen plasma.

Samples undergoing typical high-temperature APAM processing were degassed at 550 °C for 20 min with a pBN heating element to make the silicon sufficiently conductive to easily carry a DC current (a standard APAM heating process) and then at 750 °C for 40 min via DC Joule heating. Thermal desorption of the 3 nm of oxide was accomplished through six cycles alternating between 750 °C and 950 °C–1050 °C for a total of 60 min at 950 °C–1050 °C by DC Joule heating, with temperatures measured by optical pyrometry. While higher temperatures are ordinarily used to ensure that no carbon remains on the surface, there was no evidence of carbon contamination during subsequent STM imaging. This results from assuring photoresist is only baked in contact with thermally oxidized silicon during microfabrication and using an *ex situ* oxygen plasma treatment before thermal oxide desorption.

Sputtered-and-annealed samples, including blank chips for subsequent mesa-etched Hall bars (Sec. III), hardmask-defined Hall bars (Sec. IV), and FEOL CMOS chips (Sec. V), were degassed at approximately 550 °C for 20 min with a heating element and then approximately 600 °C for 40 min via DC Joule heating. Sputtering to remove 3 nm of oxide was conducted by pressurizing the prep chamber to $2–5 \times 10^{-3}$ mBar neon with a 1.5 kV accelerating voltage (unless otherwise specified) applied to the magnetron-generated neon plasma to bombard the sample surface at 60° from normal incidence, aimed parallel to the narrow (5 mm) direction of the chip. No additional heat was applied to the samples during sputtering. The sputtered samples were then annealed at 300 °C with a heating element for 15 min, followed by 15 s of flash annealing through DC Joule heating at 550–600 °C, unless times or temperatures are otherwise specified.

From there, the UHV process flows of all samples reconverged. Next, the doping process was conducted by dosing the samples with phosphine for 20 min at a chamber pressure of $1 \times 10^{-8}$ mBar. The samples were then annealed at 300 °C for 15 min to incorporate the P atoms, followed by depositing 2 nm of silicon without intentionally heating the substrate to follow a locking layer process.[19,20] This initial layer locks the phosphorus in place by minimizing segregation to the surface through adatom mediated diffusion and confining most of the phosphorus to these 2 nm of silicon during subsequent recrystallization. Thus, with most of the phosphorus buried under silicon, subsequent epitaxy may proceed at higher temperature to grow higher quality silicon with less phosphorus diffusion than would otherwise be attainable. The locking layer was recrystallized by flash annealing at 550–600 °C for 15 s by DC Joule heating. Finally, samples were held at $\sim$300 °C to grow a silicon cap via MBE at a rate of 0.5 nm/min. BEOL fabrication of APAM test devices and CMOS chips are described in Secs. VII C and VII D, respectively.

### C. Post-APAM microfabrication

After APAM processing, the samples were returned to the cleanroom. To define mesa-etched Hall bars, a $CF_4$/Ar ICP was used for a 100 nm mesa etch through the epitaxial silicon, delta layer, and into the substrate. This was followed by a 50 nm via with a $CF_4$/Ar ICP to reach through the delta layer. Metal contacts were then added using a liftoff process with an electron-beam deposited metal layer (aluminum). Hardmask samples underwent a similar BEOL process, with the exception of forgoing the mesa etch.

### D. CMOS devices

CMOS wafers were fabricated at Sandia National Laboratories using a foundry process based on a 0.35 $\mu$m gate length, 3.3 V, 7 nm gate oxide, shallow trench isolated, partially-depleted silicon on insulator technology. A custom mask set was developed with individually-testable transistors, custom layout cells to integrate the APAM regions and associated body contacts, as well as simple integrated circuits. For this work, the discrete transistors were constructed with 3 $\mu$m gate length and 13 $\mu$m gate width, and the APAM regions had patterned implants and body contacts compatible with APAM integration. The FEOL proceeded with standard fab processes for ion implantation and activation. The MOL consisted of self-aligned titanium silicide formation, where pre-APAM regions were protected from silicidation using a silicon nitride cap, followed by deposition of high-density plasma oxide + nitride pre-metal dielectric (PMD) approximately 1 $\mu$m thick, and finally damascene tungsten contact plugs deposited by chemical vapor deposition, along with a Ti/TiN liner deposited by physical vapor deposition.

After contact formation, the APAM integration on CMOS chips was done by etching a $2 \times 14\ \mu m^2$ window through the PMD to expose the underlying silicon using a photoresist mask and dry etching based on $CHF_3$ chemistry. The window etch was soon followed by a thin, 3 nm PECVD oxide for protection during sample transfer. The APAM part of MOL processing then proceeded as described above in Sec. VII B. After APAM, the epitaxial silicon was etched away from the field leaving a $10 \times 18\ \mu m^2$ patch surrounding the doped APAM window. A simple BEOL metal routing layer for electrical testing was then patterned using photolithography and liftoff of an electron-beam deposited aluminum metal layer. Electrical testing was performed with a probe station for initial room-temperature testing, then using wire-bonding for integration and temperature dependent measurements.

### E. CMOS thermal budget tests

CMOS chips prepared for thermal budget testing were processed through FEOL to the point after forming the well implants and tungsten plugs for contacts. Next, the chips underwent rapid thermal annealing for 20 min at temperatures ranging from 400 °C to 900 °C. These devices then went through BEOL metal routing and bond pads. The threshold voltage of the PMOS and NMOS transistors on the chip was defined by measuring the gate bias required to reach a predetermined subthreshold drive current density under constant source-drain bias of 3.3 V. The change in threshold voltage for each sample was then determined by comparing the annealed threshold voltage to that of standard chips that were not annealed.

### F. Cryogenic device measurement

Hall measurements were performed by submerging the samples in a liquid helium Dewar to keep the device at 4.2 K. The magnetic field was swept from −1 T to 1 T with an electromagnet with a known current to field ratio. The current applied to the device and the resulting voltages were measured with three lock-in amplifiers.

Samples used to determine the thermal budget of APAM delta layers after silicon epitaxy were prepared through the high-temperature oxide desorption process described in Sec. VII B. These samples then underwent rapid thermal annealing for 15 min at temperatures ranging from 400 to 700 °C before completing the process described above in Sec. VII C. The electronic thickness of delta layers is assessed by analysis of weak localization signals under parallel and perpendicular magnetic fields using a closed-cycle pumped-helium cryostat. Data for both field orientations are collected on a single cooling cycle using an *in situ* rotator on which the sample is mounted, and the data are fitted to the Hikami−Larkin−Nagaoka equation by which the delta layer thickness is extracted.[24]

### G. Device cross section

Cross sections of samples that were examined in a Scanning Transmission Electron Microscope (STEM) were produced by focused ion beam milling with Ga ions to cut and lift out lamellae for STEM imaging. A high-angle annular dark-field (HAADF) detector was used for STEM imaging.

## ACKNOWLEDGMENTS

We would like to thank Rick Muller, Paul Sharps, Conrad James, Robert Koudelka, and David White for their management support and Mark Gunter, Phillip Gamache, Ashlyn Vigil, Alisha Hawkins, Christopher Bishop, Andrew Starcbuck, Brian Tierney, and Reza Argahvani for their technical help and advice.

This work was supported by the Laboratory Directed Research and Development Program at Sandia National Laboratories under Project No. 213017 and by the U.S. Department of Energy (DOE) Advanced Materials and Manufacturing Technologies Office (AMMTO) project BEATS. Work was performed, in part, at the Center for Integrated Nanotechnologies, a U.S. DOE, Office of Basic Energy Sciences user facility. This article has been authored by an employee of National Technology and Engineering Solutions of Sandia, LLC under Contract No. DE-NA0003525 with the DOE. The employee owns all rights, title, and interest in and to the article and is solely responsible for its contents. The U.S. Government retains, and the publisher, by accepting the article for publication, acknowledges that the U.S. Government retains a nonexclusive, paid-up, irrevocable, worldwide license to publish or reproduce the published form of this manuscript or allow others to do so, for U.S. Government purposes. The Department of Energy will provide public access to these results of federally sponsored research in accordance with the DOE Public Access Plan https://www.energy.gov/doe-public-access-plan. This paper describes objective technical results and analysis. Any subjective views or opinions that might be expressed in the paper do not necessarily represent the views of the U.S. Department of Energy or the U.S. Government.

## AUTHOR DECLARATIONS

### Conflict of Interest

The authors have no conflicts to disclose.

### Author Contributions

**E. M. Anderson:** Data curation (equal); Formal analysis (equal); Investigation (equal); Writing – original draft (equal); Writing – review & editing (equal). **C. Arose:** Data curation (equal); Investigation (equal). **T.-M. Lu:** Data curation (equal); Investigation (equal);

Writing – review & editing (equal). **C. Halsey:** Data curation (equal); Formal analysis (equal); Investigation (equal). **T. D. England:** Methodology (equal); Resources (equal). **D. R. Ward:** Methodology (equal); Project administration (equal); Resources (equal); Writing – review & editing (equal). **D. A. Scrymgeour:** Project administration (equal); Writing – review & editing (equal). **S. Misra:** Conceptualization (equal); Funding acquisition (equal); Project administration (equal); Writing – original draft (equal); Writing – review & editing (equal). **C. R. Allemang:** Data curation (equal); Formal analysis (equal); Investigation (equal); Writing – original draft (equal); Writing – review & editing (equal). **A. J. Leenheer:** Data curation (equal); Formal analysis (equal); Investigation (equal); Writing – original draft (equal); Writing – review & editing (equal). **S. W. Schmucker:** Data curation (equal); Formal analysis (equal); Investigation (equal); Writing – original draft (equal); Writing – review & editing (equal). **J. A. Ivie:** Data curation (equal); Formal analysis (equal); Investigation (equal); Writing – original draft (equal); Writing – review & editing (equal). **D. M. Campbell:** Data curation (equal); Formal analysis (equal); Investigation (equal); Writing – review & editing (equal). **W. Lepkowski:** Formal analysis (equal); Investigation (equal). **X. Gao:** Data curation (equal); Formal analysis (equal); Investigation (equal); Writing – original draft (equal); Writing – review & editing (equal). **P. Lu:** Data curation (equal); Investigation (equal).

## DATA AVAILABILITY

The data that support the findings of this study are available within the article.